\documentclass{optica-article}

\journal{opticajournal} 

\articletype{Research Article}

\usepackage{lineno}
\usepackage{siunitx}
\usepackage{booktabs}
\usepackage{multirow}
\usepackage{blindtext}
\usepackage{comment}

\usepackage{soul} 
\usepackage{xcolor}

\begin{document}

\title{Cordierite-based optical resonators with extremely low thermal expansion}

\author{Nico Wagner,\authormark{1,2,$\dag$} Thomas Legero,\authormark{3,$\dag$,*} and Stefanie Kroker\authormark{1,2,3}}

\address{
\authormark{1}Institut für Halbleitertechnik, Technische Universität Braunschweig, Hans-Sommer-Str. 66, 38106 Braunschweig, Germany\\
\authormark{2}Laboratory for Emerging Nanometrology, Langer Kamp 6a-b, 38106 Braunschweig, Germany\\
\authormark{3}Physikalisch-Technische Bundesanstalt, Bundesallee 100, 38116 Braunschweig, Germany\\
\authormark{$\dag$}The authors contributed equally to this work.
}

\email{\authormark{*}thomas.legero@ptb.de}

\begin{abstract*}
Applications for ultra-stable lasers outside controlled laboratory environments require compact and robust optical resonators with reduced sensitivity to temperature fluctuations.
The low thermal expansion coefficient (CTE) and the high stiffness make cordierite-based ceramics, such as NEXCERA\textsuperscript{\texttrademark}, attractive for vibration insensitive room-temperature resonators.
We revisit the effective CTE of resonators with spacers and mirrors made of different materials and use finite element simulations to analyze the impact of a CTE mismatch in a cordierite-based resonator with mirrors made of ultra-low expansion (ULE\textsuperscript{\textregistered}) glass or fused silica (FS).
This enabled us to determine the CTE of a cordierite spacer from the measured effective CTE of a resonator.
We confirm a six-fold larger CTE slope of cordierite around the zero-crossing temperature than in ULE glass. 
The steep CTE slope, in combination with the large stiffness, makes cordierite-based resonators far less sensitive to CTE mismatch with FS mirrors, thereby eliminating the need for additional compensation rings.
We further consider the so far neglected case, where the CTE of the spacer is larger than that of the mirror, and propose resonator designs in which the thermal length change of the spacer is fully or partially compensated by the deflection of the mirrors.
This results in a cordierite-based resonator with ULE mirrors whose effective CTE can be close to zero over a temperature range of several tens of Kelvin.
We are extending our concept to resonators based on crystalline materials with high stiffness and low isothermal length change, such as silicon, enabling compact and robust room-temperature resonators for terrestrial and space-born applications.
\end{abstract*}

\section{Introduction}

Fabry-Pérot optical resonators are essential instruments in high-precision measurement science, with applications in optical lattice clocks~\cite{Ludlow_2015, Matei_2017, Zhang_2017, Oelker2019}, gravitational-wave detection~\cite{Punturo2010, Aso_2013, Acernese2014, LIGO2015, Abac2025}, fundamental physics experiments~\cite{Ashby2003, Roberts2017, Delva2018, Milner2019, Kennedy2020} and exoplanet detection~\cite{Kreider2025}.

The performance of such systems is limited by the resonators length stability, requiring materials with an extremely low or near-zero coefficient of thermal expansion (CTE).
A prominent example is Corning’s ultra-low expansion (ULE\textsuperscript{\textregistered}) glass, whose CTE has a zero crossing at room temperature.
However, ULE exhibits a relatively low Young's modulus and show large isothermal length drift due to material aging~\cite{Phillips_1996, Dube_2009}.

Cordierite-based ceramics, such as NEXCERA\textsuperscript{\texttrademark} by Krosaki Harima Corporation, are considered as a promising alternative~\cite{Barbarat2025}. 
Its CTE also shows a zero crossing at room temperature and there are hints that its isothermal length drift is smaller than in ULE glass~\cite{Ito2017}.
The Young's modulus (\SI{145}{\giga\pascal}~\cite{NipponSteel2001}) is more than a factor of two larger than ULE glass (\SI{67.6}{\giga\pascal}~\cite{ULE_7972}) which makes the resonator length less susceptible to acceleration, resulting in lower vibration sensitivity.
Although the mechanical loss factor of NEXCERA is comparable to that of ULE~\cite{Wagner2025_NEXCERA, Wagner2025_SPIE}, its higher Young’s modulus leads to reduced Brownian thermal noise.

Aside from ultra-stable laboratory systems, there is growing demand for compact optical resonators that can be operated outside of controlled laboratory environments.
Applications range from photonic radar systems~\cite{Pan2020} and environmental sensing~\cite{Ip2022} to mobile atomic clocks~\cite{Cacciapuoti2009}.
Recent developments in microcavity technologies further highlight this trend toward miniaturization, enabling resonators with small mode volumes and high finesse for applications in quantum technologies and integrated photonics~\cite{Ding2026}.
However, the absence of controlled laboratory conditions, in particular lower temperature stability and higher vibration noise, place special demands on the CTE of the resonator and the mechanical properties of its materials.
In this article, we show that cordierite, in addition to its excellent mechanical properties, is also an ideal material for such applications in terms of its CTE.

The paper is organized as follows.
In Sec.~\ref{sec:thermal_expansion} we revisit the thermal expansion of optical resonators composed of different materials.
This leads us to design strategies for realizing resonators with extremely small effective CTEs.
For example, cordierite can be used as a spacer material in order to benefit from higher rigidity, lower isothermal drift and significantly reduced sensitivity to temperature fluctuations.

Section~\ref{sec:cordierite_delta} revisits the impact of a large CTE mismatch between spacer and mirror material. 
This is particularly important when using fused silica (FS) mirrors which is a suitable choice due to its low thermal noise at room temperature~\cite{Penn2001}. 
As shown in \cite{Legero_2010}, ULE resonators employing FS mirrors suffer from significant mirror deformations, leading to an effective CTE of the resonator that can differ substantially from that of the spacer material.
Using finite-element (FEM) simulations, we show that the large slope of cordierites CTE and its high Young's modulus strongly reduce this effect, eliminating the need for ULE rings.

Designing resonators with the lowest possible effective CTE requires a good knowledge of the CTE parameters of the spacer and mirror materials. 
We have measured the effective CTE of a test resonator consisting of a cordierite spacer and ULE mirrors with known CTE. 
Based on our analysis and the FEM simulation, we determined the CTE of the cordierite spacer.
Our results presented in Sec.~\ref{sec:experiment} confirm that the CTE of cordierite exhibits a steep slope of \SI{8.811+-0.025e-9}{\per\K\squared} at its zero-crossing temperature.

In Sec.~\ref{sec:hybrid_designs}, we propose a resonator design that combines a cordierite spacer with ULE mirrors, whose effective CTE remains nearly zero over several Kelvin. 
This concept can be realized with compact cavities in the \SI{1}{\centi\meter} range. 
We further extend this approach by introducing two additional concepts. 
One concept considers resonators whose effective CTE is determined solely by the mirror substrates, allowing the spacer to be fabricated from crystalline materials with low length drift and high Young's modulus. 
Finally, we discuss a silicon cavity with FS mirrors operated at room temperature that exhibits an effective CTE comparable to that of ULE glass, enabling a thermal-noise-limited frequency stability on the order of \num{1e-15}. 
All designs facilitate compact and transportable systems that combine excellent thermal stability with robust mechanical performance at room temperature.

\section{Thermal expansion of optical resonators}\label{sec:thermal_expansion}

The theoretical description of thermal expansion in composite materials has been discussed in \cite{Nottcut_1995, Wong_1997}.
In this work, we adopt the notation of~\cite{Legero_2010}, considering a cylindrical resonator of length $L$ with optically contacted mirrors of radius $R$.
When the resonator undergoes a temperature change, the mismatch in the CTEs between the spacer and the mirrors leads to a differential radial expansion between spacer and mirror.
This radial expansion, $\mathrm{d}R$, can be expressed as $\mathrm{d}R = (\alpha_{\mathrm{m}} - \alpha_{\mathrm{s}}) R \, \mathrm{d}T$, where $\alpha_{\mathrm{m}}$ is the CTE of the mirror and $\alpha_{\mathrm{s}}$ is the CTE of the spacer.
As the mirrors are rigidly contacted to the spacer, this radial expansion is suppressed and results in a bulging of the mirrors, which leads to an axial displacement, $\mathrm{d}A$, of the mirror center.
According to the linear stress-strain relationship, the radial expansion $\mathrm{d}R$ is related to the axial displacement $\mathrm{d}A$ through a coupling coefficient $\delta$, such that $\mathrm{d}A = \delta \, \mathrm{d}R$.

The resonator length change $\mathrm{d}L$ is then given by $\mathrm{d}L = \alpha_{\mathrm{eff}} L \, \mathrm{d}T$, with an effective CTE, $\alpha_{\mathrm{eff}}$.
Since the total resonator length change includes the expansion of the spacer and twice the axial expansion of the mirrors, the following relation holds
\begin{equation}\label{eq:alpha_length_change}
    \alpha_{\mathrm{eff}} L \, \mathrm{d}T = \alpha_{\mathrm{s}} L \, \mathrm{d}T + 2 \, \mathrm{d}A  \, .
\end{equation}
Substituting $\mathrm{d}A = \delta (\alpha_{\mathrm{m}} - \alpha_{\mathrm{s}}) R \, \mathrm{d}T$ into Eq.~(\ref{eq:alpha_length_change}) yields the effective CTE as 
\begin{equation}\label{eq:effective_cte}
    \alpha_{\mathrm{eff}} (T) = \alpha_{\mathrm{s}} (T) + k \left[ \alpha_{\mathrm{m}} (T) - \alpha_{\mathrm{s}} (T) \right] \, ,
\end{equation}
with $k=2R\delta/L$ as the correction factor.
Note, that the coupling coefficient $\delta$ is only dependent on the geometry of mirror and spacer and their mechanical properties and not from the CTE of the materials.

This approach was first discussed to describe the effective CTE of ULE-based resonators with FS mirrors with a typical coupling coefficient of $\delta \approx 0.4$~\cite{Legero_2010}. 
In this case we have $\alpha_{\mathrm{m}}(T) > \alpha_{\mathrm{s}}(T)$ and because $k > 0$ the thermally induced mirror deformation has the same sign as the spacer length change. As a result the effective CTE is larger than the spacer's CTE and the zero-crossing temperature (ZCT) of the resonator can be decreased by several 10 Kelvin.
The only way to reduce this effect is to aim for a small value of $k$.
For long resonators with $R/L < 1$ the correction factor $k$ is small and this effect is reduced.
Short resonators can be equipped with compensation rings on the backside of the mirrors, suppressing mirror deformation and ideally leading to $k \approx 0$~\cite{Legero_2010}.
We will address this case in Sec.~\ref{sec:cordierite_delta} and Sec.~\ref{sec:experiment}.

Here, we propose considering the opposite scenario, in which the resonator materials are chosen such that $\alpha_{\mathrm{m}}(T) < \alpha_{\mathrm{s}}(T)$.
In this configuration, the mirror deformation counteracts the length change of the spacer, which allows us to realize resonators whose effective CTE can be even smaller than the CTE of its components.

The goal is to realize an optical resonator with a ZCT, where the sensitivity to temperature fluctuations is minimized, i.e. $\alpha_{\mathrm{eff}}(T_{\mathrm{eff}})=0$. 
Imposing this condition, Eq.~(\ref{eq:effective_cte}) can be rearranged to determine the correction factor $k$ that fulfills it. 
The correction factor is then given by
\begin{equation}\label{eq:general_k}
    k = \frac{\alpha_{\mathrm{s}}(T_{\mathrm{eff}})}
           {\alpha_{\mathrm{s}}(T_{\mathrm{eff}}) - \alpha_{\mathrm{m}}(T_{\mathrm{eff}})} .
\end{equation}
Here, $T_{\mathrm{eff}}$ is the desired effective temperature, which can be chosen so that the correct value for $k$ is determined.
Close to the ZCT, the CTE can be approximated by a linear function $\alpha_i = a_i \cdot(T - T_i)$~\cite{Legero_2010}, so that
\begin{equation}\label{eq:k_first}
    k = \frac{
        a_{\mathrm{s}} \cdot(T_{\mathrm{eff}}-T_{\mathrm{s}})}
        {a_{\mathrm{s}} \cdot(T_{\mathrm{eff}}-T_{\mathrm{s}})
        - a_{\mathrm{m}} \cdot(T_{\mathrm{eff}}-T_{\mathrm{m}})} .
\end{equation}
This equation of the correction factor depends on the desired material combination and represents the value that the resonator must have in order to achieve $\alpha_{\mathrm{eff}}(T_{\mathrm{eff}})=0$.

The achievable range of $k$ is limited due to the saturation of $\delta$ for large geometries, restricting the feasible material combinations and resonator geometries. Moreover, the resulting effective slope of the CTE around the ZCT remains a critical parameter, as it depends on the material properties of both components.

Two further special cases can be derived from Eq.~(\ref{eq:effective_cte}).  
First, consider the case $k=1$. Then, in linear approximation, $\alpha_{\mathrm{eff}}(T)=\alpha_{\mathrm{m}}(T)$, meaning that the spacer contribution to the effective CTE vanishes.
This enables the use of spacer materials with significantly higher rigidity and lower isothermal drift, e.g. silicon, sapphire or diamond. 

Second, the CTE of the spacer $\alpha_{\mathrm{s}}$ as well as the second term $k ( \alpha_{\mathrm{m}} (T) - \alpha_{\mathrm{s}} (T))$ can be compensated if $k=a_{\mathrm{s}}/(a_{\mathrm{s}}-a_{\mathrm{m}})$.
Substituting this $k$ expression into Eq.~(\ref{eq:effective_cte}) and using the linear approximation of the CTE leads to a constant effective CTE
\begin{equation}\label{eq:alpha_const}
    \alpha_{\mathrm{eff}}\left(k = \frac{a_{\mathrm{s}}}{a_{\mathrm{s}} - a_{\mathrm{m}}}\right)
    = \frac{a_{\mathrm{m}} a_{\mathrm{s}} (T_{\mathrm{m}} - T_{\mathrm{s}})}{a_{\mathrm{m}} - a_{\mathrm{s}}} .
\end{equation}
This temperature-independent CTE is close to zero if $T_{\mathrm{m}}\approx T_{\mathrm{s}}$, with a remaining contribution determined by the nonlinear components of the CTE.
We will discuss possible resonator configurations in more detail in Sec.~\ref{sec:hybrid_designs}.

\section{\label{sec:cordierite_delta}Cordierite resonators with fused silica mirrors}

We utilize FEM simulations (COMSOL\textsuperscript{\textregistered} Multiphysics~\cite{Comsol}) to evaluate the coupling coefficient $\delta$ for a cordierite spacer with FS mirror substrates. 
For the simulations, the CTEs of the spacer and mirror materials must be assumed.
However, they do not influence the resulting value of the coupling coefficient. 
The resonator deformation $\Delta L(T)$ is simulated by varying the temperature $T$, starting from the spacer ZCT $T_{\mathrm{s}}$. 
The resulting length change is described by the effective CTE as defined in Eq.~(\ref{eq:effective_cte}). 
The effective ZCT $T_{\mathrm{eff}}$ of the resonator is obtained from a least-squares fit to the simulated data and is then used to determine the coupling coefficient $\delta$ via Eq.~(\ref{eq:effective_cte}). 
The material parameters used in the simulations are listed in Tab.~\ref{tab:material_parameters}.

\begin{figure}[t]
    \centering
    \includegraphics[width=1.0\linewidth]{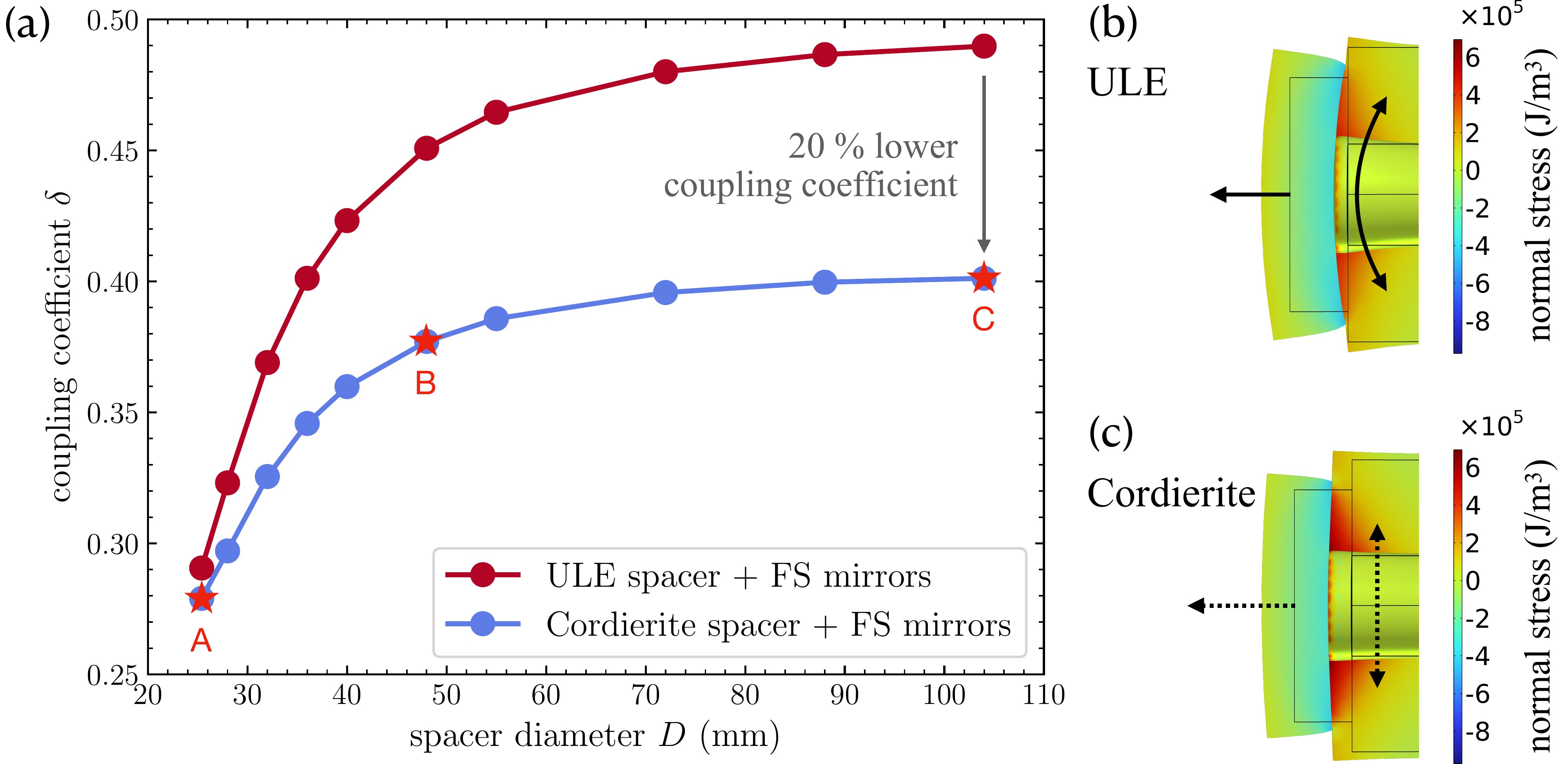}
    \caption{(a) Simulated coupling coefficient $\delta$ for a resonators with FS mirrors based on a ULE or cordierite spacer. The spacer length is \SI{105.5}{\milli\meter}, with outer and inner bore diameters of \SI{32}{\milli\meter} and \SI{11}{\milli\meter}, respectively. The mirrors have a diameter of \SI{25.4}{\milli\meter} and a thickness of \SI{6.3}{\milli\meter}. (b) and (c) show a zoom-in of the elastic deformation at the mirror–spacer interface caused by a temperature increase of \SI{20}{\K}, for the same resonator geometry with a ULE spacer and a cordierite spacer, respectively. The color scale represents the normal stress distribution within the resonator. 
    The higher stiffness of cordierite reduces the coupling coefficient because the mirror undergoes less bending under thermally induced stresses. As a result, the coupling coefficient is up to \SI{20}{\percent} lower compared to ULE-based resonators.
    The marked points correspond to Fig.~\ref{fig:delta_over_L}, where they appear as data points in the length-dependent coupling coefficient plot.}
    \label{fig:delta_over_D}
\end{figure}

\begin{table}[tb]
    \caption{Material parameters assumed for the COMSOL simulations to determine the coupling coefficient $\delta$.}
    \label{tab:material_parameters}
    \centering
    \begin{tabular}{cccc}
        \toprule
         & Cordierite~\cite{NipponSteel2001} & ULE\textsuperscript{\textregistered}~\cite{ULE_7972} & Fused Silica~\cite{FS_Corning} \\
        \hline
        $E$ (\si{\giga\pascal}) & \num{145} & \num{67.6} & \num{72.7} \\
        $\nu$ (-) & \num{0.31} & \num{0.17} & \num{0.16} \\
        $\rho$ (\si{\kg\per\meter\cubed}) & \num{2500} & \num{2210} & \num{2200} \\
        $C_p$ (\si{\J\per\kg\per\kelvin}) & \num{770} & \num{767} & \num{703} \\
        $\kappa$ (\si{\W\per\meter\per\kelvin}) & \num{4.2} & \num{1.31} & \num{1.38} \\
        \bottomrule
    \end{tabular}
\end{table}

Figure~\ref{fig:delta_over_D}(a) compares the resulting coupling coefficients of cavities with FS mirrors on an ULE and a cordierite spacer, respectively.
The coupling coefficient for the cordierite resonator is up to \SI{20}{\percent} lower for larger spacer diameters.
This difference is explained by analyzing the axial deformation of the resonator, shown in Fig.~\ref{fig:delta_over_D}(b) and (c).
Due to the higher Young's modulus of cordierite, the deformation of the mirror is more constrained resulting in less bending and thus a lower $\mathrm{d}A/\mathrm{d}R$ ratio.

\begin{figure}[tb]
    \centering
    \includegraphics[width=0.8\linewidth]{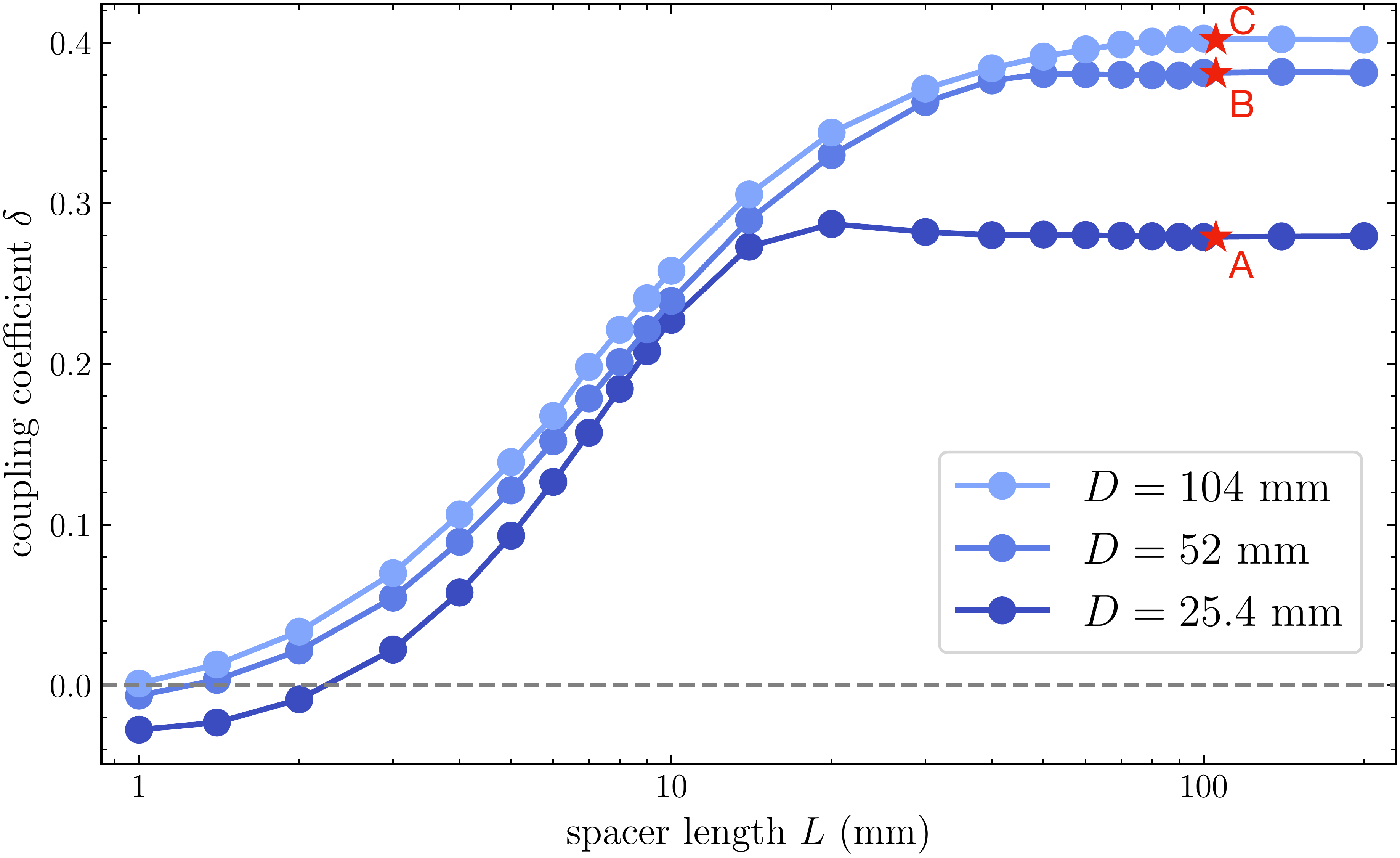}
    \caption{Influence of the spacer length $L$ on the coupling coefficient $\delta$ for spacer configurations with $D = \SI{25.4}{\milli\meter}$, $D = \SI{52}{\milli\meter}$, and $D = \SI{104}{\milli\meter}$, over a length range from \SI{1}{\milli\meter} to \SI{200}{\milli\meter}. With shorter spacers, the coupling coefficient is strongly dependent on the length. The results show that the coupling coefficient becomes constant once the spacer length is larger than its diameter. The marked points are related to Fig.~\ref{fig:delta_over_D}.}
    \label{fig:delta_over_L}
\end{figure}

While the spacer length was fixed to \SI{105.5}{\milli\meter} in our experiments, we also highlight its influence on the coupling coefficient since this becomes important for short resonators.
In Fig.~\ref{fig:delta_over_L} we present the coupling coefficient as a function of spacer length from \SI{1}{\milli\meter} to \SI{200}{\milli\meter} for fixed spacer diameters of \SI{25.4}{\milli\meter}, \SI{52}{\milli\meter}, and \SI{104}{\milli\meter}.
The coupling coefficient decreases once the spacer length is smaller than the spacer diameter.
Please note that when the spacer length exceeds its diameter, the coupling coefficient approaches a constant value.
For spacers short compared to the mirror volume, the mirror deformations are not independent from each other and dominate the deformation of the entire resonator.
In our configuration, this is achieved for lengths below \SI{10}{\milli\meter}, where the coupling coefficient decreases drastically and reaches zero or even negative values.
Furthermore, since the coupling coefficient $\delta$ is proportional to the axial expansion mismatch $\mathrm{d}A$, a negative $\delta$ observed for short resonators indicates that local mirror bending dominates over the spacer’s contribution.
It is also important to note that $\delta$ is bound to a maximum value of \num{0.4} given by the mechanical properties of the mirror and the spacer.

\section{\label{sec:experiment}Experimental determination of the CTE of cordierite}

First measurements on the CTE of cordierite spacers with ULE mirrors have been reported in \cite{Hosaka_2013} and \cite{Kwong_2018}. In these experiments, the CTE mismatch between spacer and mirror material and the coupling coefficient $\delta$ are not considered and the measurement results do not represent the actual CTE of cordierite.

Our simulation results of $\delta$ allows us to estimate the impact of a CTE mismatch between spacer and mirrors and to calculate the CTE of the spacer material from equation~(\ref{eq:effective_cte}). For this, we have measured the effective CTE of a resonator made of a \SI{100}{\milli\meter} long cordierite spacer and ULE mirrors with known CTE.
The CTE of the mirror pair was measured using a spacer made from the same piece of ULE glass as the mirrors.

\begin{figure}[t]
    \centering
    \includegraphics[width=1.0\linewidth]{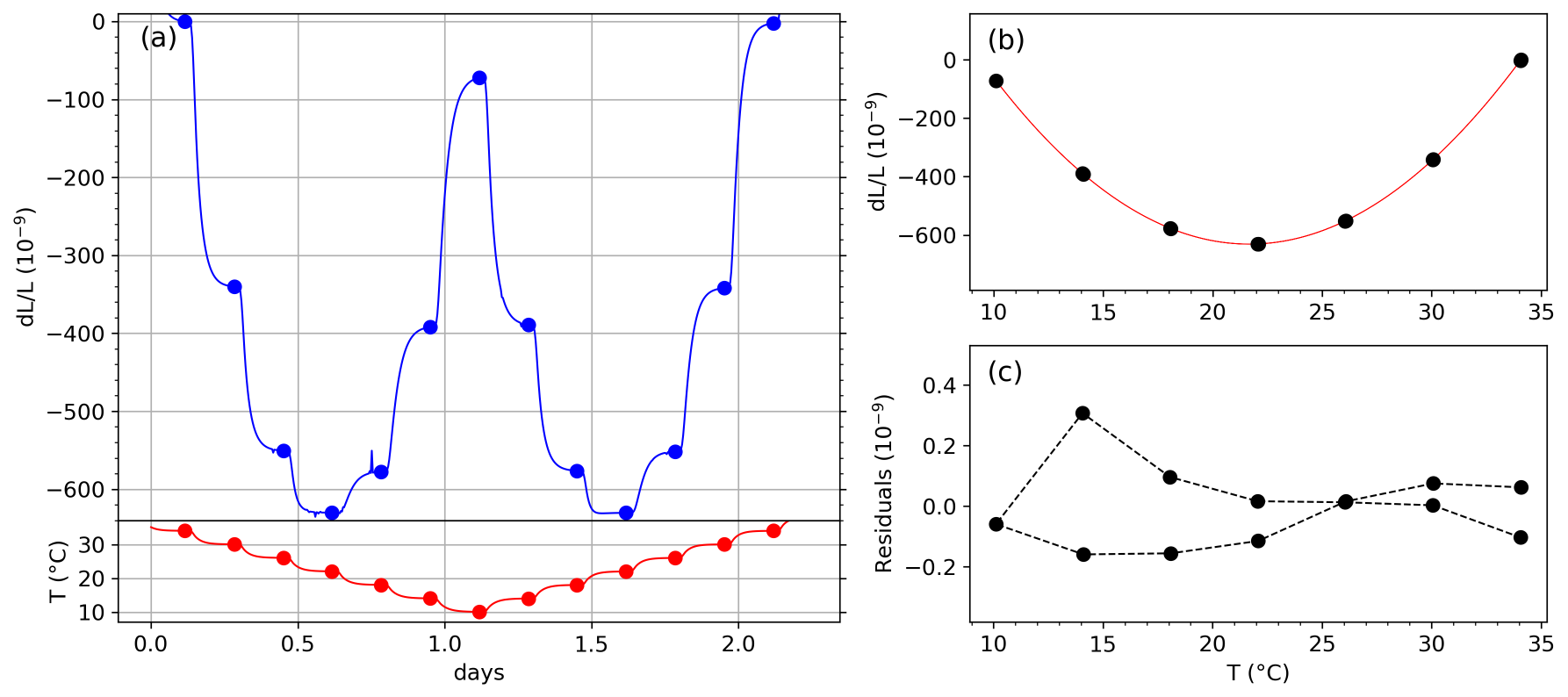}
    \caption{(a) Relative length change of the resonator during a stepwise temperature ramp. The \SI{4}{\K} temperature steps are separated by a thermalization time of 4 hours. The blue and red dots mark the achieved length change and temperature at the end of each step. (b) Temperature-dependent length changes. The black dots are the measurement points from (a). A least squares fit of Eq.~(\ref{eq:length_change}) is shown as red line. (c) The residuals of the fit are well below \SI{0.1}{\percent}.}
    \label{fig:cte_measurement}
\end{figure}

The cordierite spacer and the ULE mirrors have a diameter of \SI{25.4}{\milli\meter}. 
The spacer has a bore of \SI{10}{\milli\meter} and a centered venting hole of \SI{4}{\milli\meter}. 
The \SI{6}{\milli\meter} thick, concave ULE mirrors have a radius of curvature of \SI{150}{\milli\meter}.
A contacting annulus of about \SI{7}{\milli\meter} allows for optical contacting to the spacer.
The mirrors are equipped with a dielectric coating with a reflectivity of \SI{99.85}{\percent} at a center wavelength of \SI{633}{\nano\meter}. 

Our measurement follows the technique described in \cite{Legero_2010}. 
The resonator is placed in a temperature-controlled enclosure in a vacuum chamber with a residual pressure of approximately \SI{e-5}{\milli\bar}. 
Its temperature is measured on the spacer surface using a calibrated PT-100.  

We have used a lock-in technique to stabilize the frequency $\nu$ of a He-Ne laser at \SI{633}{\nano\meter} to a TEM$_{00}$ mode of the resonator. 
The relative length change is calculated from the relative frequency change of the resonator by $\mathrm{d}L/L = - \mathrm{d}\nu/\nu$.
We determine the frequency $\nu$ of the He-Ne laser as well as the frequency changes using the beat signal with a \SI{633}{\nano\meter} reference laser. 

During the measurement, the temperature of the enclosure is changed in steps of \SI{4}{\kelvin}. 
The thermal time constant of the resonator is approximately 45 min. 
A thermalization time of 4 hours was therefore allowed between the steps.
The results are shown in Fig.~\ref{fig:cte_measurement}(a).
The temperature steps followed a ramp from \SI{34}{\degreeCelsius} to \SI{10}{\degreeCelsius} and back. 
The total measurement time was approximately 2 days. 
We select the temperature values and corresponding relative length changes at the end of each step to obtain $\mathrm{d}L/L$ as a function of $T$.

The result is shown in Fig.~\ref{fig:cte_measurement}(b).
The red line represents a least squares fit to the measured data.
For this we have approximated the CTE by a second-degree polynomial
\begin{equation}\label{eq:cte-model}
\alpha_{\mathrm{eff}} (T) = a_{\mathrm{eff}} \cdot \left( T - T_{\mathrm{eff}} \right) + b_{\mathrm{eff}} \cdot \left( T - T_{\mathrm{eff}} \right)^2,
\end{equation}
leading to a relative length change of
\begin{equation}\label{eq:length_change}
    \dfrac{\mathrm{d}L}{L} (T) = C + \dfrac{a_{\mathrm{eff}}}{2} \cdot \left( T - T_{\mathrm{eff}} \right)^2 + \dfrac{b_{\mathrm{eff}}}{3} \cdot \left( T - T_{\mathrm{eff}} \right)^3 + d \cdot (t-t_0)
\end{equation}
with an integration constant $C$. Here, we include a linear drift rate $d$ to take an isothermal fractional length drift of the resonator into account. 

The resulting CTE parameters of the resonator are listed in Tab.~\ref{tab:cte_results}. 
The given standard uncertainties include the fit uncertainties and systematic uncertainties of our measurement.
The measured slope $a_{\mathrm{eff}}$ of the resonators CTE is similar to the slope of \SI{7.58e-9}{\per\kelvin\squared} reported in \cite{Kwong_2018}.
Please note that this value is more than six times larger than the CTE slope of ULE glass.

For the drift rate we obtained $d = \SI{-1.6e-14}{\per\second}$.
This value is a factor 30 larger than the drift rate measured in \cite{Kwong_2018}, but we do also observe, that the drift rate of such resonators starts with large values after optical contacting and decays during the time of operation in vacuum. 

After the measurement, the mirrors were de-contacted from the cordierite spacer and optically contacted to the ULE spacer.
Apart from the length, which is \SI{56}{\milli\meter}, this spacer has the same dimensions as the cordierite sample.
The CTE parameters of this resonator were measured the same way as described above.
Since the mirror and the spacer are made of the same material, the CTE mismatch is zero and the obtained results represent the CTE parameters of the mirror pair. 
Table~\ref{tab:cte_results} summarizes the parameters of $\alpha_{\mathrm{m}}$.

\begin{table}[tb]
    \caption{CTE parameters of the resonator $\alpha_{\mathrm{eff}}$, the used mirror pair $\alpha_{\mathrm{m}}$ and the calculated CTE of the spacer material $\alpha_{\mathrm{s}}$ after correcting the CTE mismatch between mirror and spacer.}
    \label{tab:cte_results}
    \centering
    \begin{tabular}{cccc}
    \toprule
     &  resonator ($\alpha_{\mathrm{eff}}$) & mirror ($\alpha_{\mathrm{m}}$)  &  spacer ($\alpha_{\mathrm{s}}$) \\
    \hline
    $T_{0}$ (\SI{}{\degreeCelsius}) & \num{21.67+-0.16} & \num{28.97+-0.16} & \num{21.58+-0.16} \\
    $a$ (\SI{e-9}{\per\kelvin\squared}) & \num{8.2838+-0.0072} & \num{1.4171+-0.0043} & \num{8.811+-0.025} \\
    $b$ (\SI{e-12}{\per\kelvin\cubed}) & \num{-10.00+-0.38} & \num{-10.37+-0.98} & \num{-9.97+-0.42} \\
    \bottomrule
    \end{tabular}
\end{table}

The coupling coefficient and thus the correction factor $k$ was determined by FEM simulations as described in Sec.~\ref{sec:cordierite_delta}.
With the material parameters listed in Tab.~\ref{tab:material_parameters} we determine a coupling coefficient of $\delta = 0.286 \pm 0.001$, which gives a correction factor of $k=2R\delta/L = 0.073 \pm 0.004$.
The uncertainties are derived from worst case estimates using the machining tolerances of mirrors and spacer of \SI{0.1}{\milli\meter}.

Equation~(\ref{eq:effective_cte}) is then used to calculate the CTE of the spacer. 
For this we assume a quadratic approximation for $\alpha_{\mathrm{s}}$ and $\alpha_{\mathrm{m}}$ analogue to Eq.~(\ref{eq:cte-model}).
Equating the coefficients of Eq.~(\ref{eq:effective_cte}) in respect to $T$ leads to a set of three equations allowing us to calculate $a_{\mathrm{s}}$, $b_{\mathrm{s}}$ and $T_{\mathrm{s}}$ of the spacer as listed in Tab.~\ref{tab:cte_results}. 

The CTE parameters of the spacer are slightly different from the measured values of the resonator.
The most important difference arise in the CTE slope.
In our cordierite spacer, we get a final value of $a_{\mathrm{s}} = \SI{8.811+-0.025e-9}{\per\kelvin\squared}$.
The difference is important because a good knowledge of the CTE parameters of the spacer and the mirror material is required to determine the necessary correction factor $k$ for a resonator with an extremely small CTE according to Eq.~(\ref{eq:k_first}).

Please note, that the quadratic parameter $b_{\mathrm{s}}$ of cordierite is very close to the quadratic parameter of ULE. 
This is an important insight for the realization of a resonator with a negligible CTE according to equation 5. With the calculated required correction factor $k$, the linear terms of the CTE cancel out, so that the temperature dependence of the effective CTE is given by the difference of the quadratic CTE parameters. We will elaborate on this in more detail in Sec.~\ref{subsec:a_const}.

Finally, we would like to stress that cordierite based resonators are less sensitive to a mismatch in the ZCT of mirrors and spacer. 
We already see this in our measurement, in which the ZCTs of spacer and mirrors differ by more than \SI{7}{\kelvin} while the effective ZCT is reduced by less than \SI{100}{\milli\kelvin}.
One reason is the smaller coupling coefficient $\delta$ due to the high Young's modulus of cordierite.
An even more significant factor is the large slope of the CTE.
Since the ZCT shift is proportional to the ratio between the CTE offset induced by the mirror substrates and the intrinsic CTE slope of the spacer material ($\Delta T_{\mathrm{ZCT}}\sim\Delta\alpha/a$), a large slope results in a smaller temperature shift of the zero crossing.
This becomes even more obvious for resonators employing FS mirrors. 
The typical ZCT shifts of resonators using ULE spacers and FS mirrors are on the order of \SI{-20}{\kelvin}, depending on the resonator length and diameter.
The reason for this is the similar slope of the CTE of both materials, while the ZCTs differ by more than \SI{100}{\kelvin}~\cite{Jacobs1981}. 
In such resonators the ZCT shift can be compensated by ULE rings on the back of the mirrors~\cite{Legero_2010}, which zero the correction factor $k$ and minimize the impact of the CTE mismatch.

In contrast, the large CTE slope of cordierite significantly reduces the impact of the CTE mismatch.
Our simulations show a maximum ZCT shift of only about \SI{-3}{\kelvin}, even when the spacer diameter is increased to \SI{104}{\milli\meter}.
Compensation rings are therefore dispensable for cordierite-based resonators, an important practical advantage of cordierite spacers compared to ULE.

\section{\label{sec:hybrid_designs}Resonators with extremely low CTE}

In Sec.~\ref{sec:thermal_expansion}, we proposed designing resonators such that the mirror deformation and the spacer’s length change counteract each other. Here, we would like to present three possible resonator configurations based on this approach.

\subsection{\label{subsec:a_const}Zero-CTE resonators}
The most interesting application of our idea relies on Eq.~(\ref{eq:alpha_const}) involving cordierite-based resonators with ULE mirrors. 
Based on the measured CTE slopes given in Tab.~\ref{tab:material_parameters}, we calculate a required correction factor of $k=\num{1.19}$.
The effective CTE of such a resonator then depends mainly on the difference between the ZCTs of the two materials.

\begin{figure}[t]
    \centering
    \includegraphics[width=0.9\linewidth]{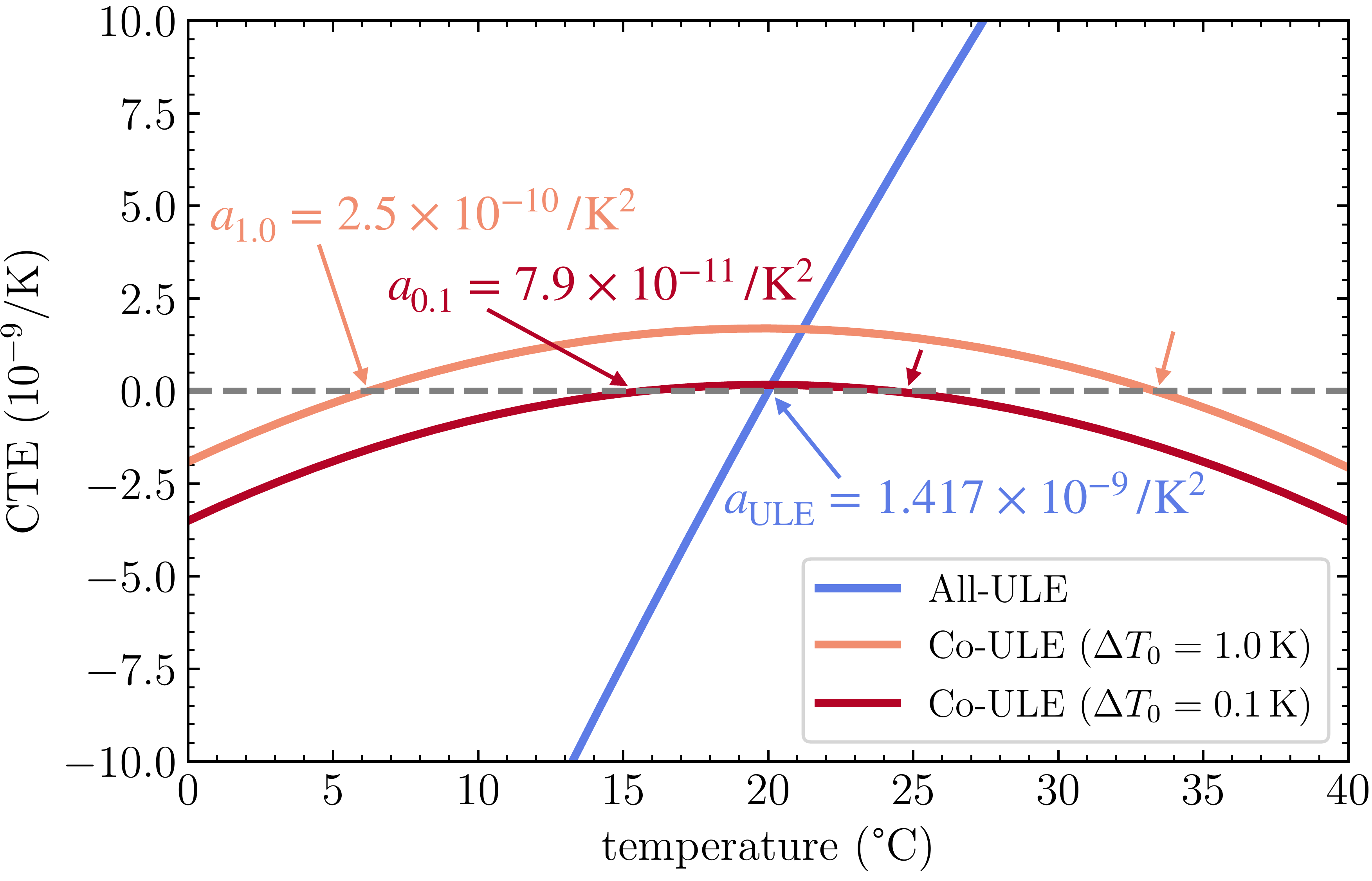}
    \caption{Calculated effective CTEs using Eq.~(\ref{eq:cte-model}) and the linear and quadratic coefficients from Tab.~\ref{tab:cte_results}. Shown is a resonator with a cordierite (Co) spacer and ULE mirrors for ZCT differences of \SI{0.1}{\K} and \SI{1}{\K} between spacer and mirror materials, with a correction factor of $k=\num{1.19}$. For comparison, the CTE of an all-ULE resonator is included. For a \SI{0.1}{\K} ZCT difference, the resonator’s CTE exhibits a slope of only \SI{7.9e-11}{\per\K\squared}, which is 18 times lower than that of ULE.}
    \label{fig:CTE_T_aeff=0}
\end{figure}

Since the ZCTs of cordierite and ULE glass can be controlled by the manufacturer, it should be possible to select both materials so that their ZCTs match to within one Kelvin or less.
Our concept compensates the effect of the linear contribution of the CTE, and thus, the nonlinear contribution remains.
In Fig.~\ref{fig:CTE_T_aeff=0}, we show the effective CTE calculated using Eq.~(\ref{eq:cte-model}) for cordierite--ULE resonators with ZCT differences of \SI{0.1}{\K} and \SI{1}{\K}.

In the design considered here, two ZCTs emerge.
At these operating points, the slope of the CTE is reduced to \SI{7.9e-11}{\per\K\squared}, which is 18 times lower than that of an all-ULE resonator.
Assuming an admissible effective CTE of $|\alpha_{\mathrm{eff}}| \leq \SI{0.2e-9}{\per\kelvin}$, the resonator would tolerate temperature variations of approximately $\pm\SI{6.3}{\K}$ around \SI{20}{\degreeCelsius}, whereas the all-ULE resonator would be limited to only $\pm\SI{0.14}{\K}$ around \SI{20}{\degreeCelsius}.

An example geometry to achieve $k=\num{1.19}$ is given by a spacer length of \SI{13}{\milli\meter}, an outer diameter of \SI{50}{\milli\meter}, an inner bore of \SI{17.4}{\milli\meter}, and one-inch mirrors with a thickness of \SI{5}{\milli\meter}.
Since the correction factor remains unchanged when the resonator geometry is scaled uniformly, the use of 2-inch mirrors allows the resonator to be scaled to a length of \SI{26}{\milli\meter}.

Small variations in the correction factor $k$ lead to a horizontal shift of the effective CTE, i.e. the ZCTs are shifted while the slope of the CTE at the ZCT remains nearly unchanged. 
Furthermore, small variations in the geometrical parameters used in the FEM simulations do not lead to significant changes in the correction factor, indicating that the design is relatively robust against small geometric deviations.
For example, a \SI{1}{\percent} change of the length of the spacer results in a \SI{0.5}{\percent} change of $k$.

The frequency stability of the proposed \SI{13}{\milli\meter}-long cordierite–ULE resonator is limited by the Brownian thermal noise of the mirrors~\cite{numata2004thermal}. 
The contributions of the ULE substrate, with its relatively large loss angle ($\phi_{\mathrm{ULE}}=\num{1.67e-5}$)~\cite{numata2004thermal}, and the contribution of the coating due to the small mode diameter resulting from the short resonator length are roughly equal.
For conventional SiO$_2$/Ta$_2$O$_5$ coatings at $\lambda=\SI{1550}{\nano\meter}$ ($\phi_{\mathrm{SiO_2/Ta_2O_5}}=\num{4e-4}$)~\cite{Robinson_2021} we calculate a fractional frequency instability of $\mathrm{mod}\,\sigma_y = \num{5.5e-15}$. 
For this we assumed a mirror configuration with a concave mirror having a radius of curvature of $\SI{1}{\meter}$ and a plane mirror.
Using crystalline AlGaAs coatings with a smaller loss angle of $\phi_{\mathrm{AlGaAs/GaAs}}=\num{2.5e-5}$~\cite{cole2013tenfold} results in a slightly better instability of $\mathrm{mod}\,\sigma_y = \num{4.2e-15}$.
These frequency instabilities are sufficient for many applications, such as photonic radar systems or environmental sensing.

This design represents a substantial advancement for applications in which temperature stability is critical. Vacuum-gap resonators have recently been introduced to eliminate the need for a vacuum chamber, enabling highly compact cavity systems~\cite{Liu2024, McLemore2024}. However, because these resonators operate in ambient conditions, they exhibit increased sensitivity to temperature fluctuations of the surrounding environment. The design approach presented here directly addresses this limitation by significantly reducing the temperature sensitivity of the resonator. Consequently, combining a cordierite spacer with ULE mirrors provides a promising route to improve the thermal stability of compact vacuum-gap cavities.

\subsection{\label{subsec:CTE_mirror}Resonators with mirror-determined CTE}

As discussed in Sec.~\ref{sec:thermal_expansion}, designing a resonator with a correction factor $k=1$ eliminates the spacer’s contribution to the effective CTE. In this case, the resonator inherits the CTE of the mirror substrate. Therefore, substrates with exceptionally low CTE, such as ULE, are particularly suitable. For the spacer, crystalline materials such as silicon become attractive due to their high stiffness and exceptionally small long-term drift~\cite{Lee2026}.

To our knowledge, resonators with $k=1$ or larger have not yet been realized.
Achieving $k=1$ requires a relatively large $R/L$ ratio for typical coupling coefficients $\delta$ in the range of \num{0.3} to \num{0.4}.
Because $\delta$ itself depends on the resonator geometry and length (see Fig.~\ref{fig:delta_over_L}), the geometry corresponding to $k=1$ must be determined using FEM simulations.

As an example, a silicon resonator with ULE mirrors fulfilling $k=1$ would require a length of \SI{20}{\milli\meter}, an outer diameter of \SI{50}{\milli\meter}, and an inner bore of \SI{17.4}{\milli\meter} with one-inch ULE mirrors with a thickness of \SI{5}{\milli\meter}.

\begin{figure}[t]
    \centering
    \includegraphics[width=1.0\linewidth]{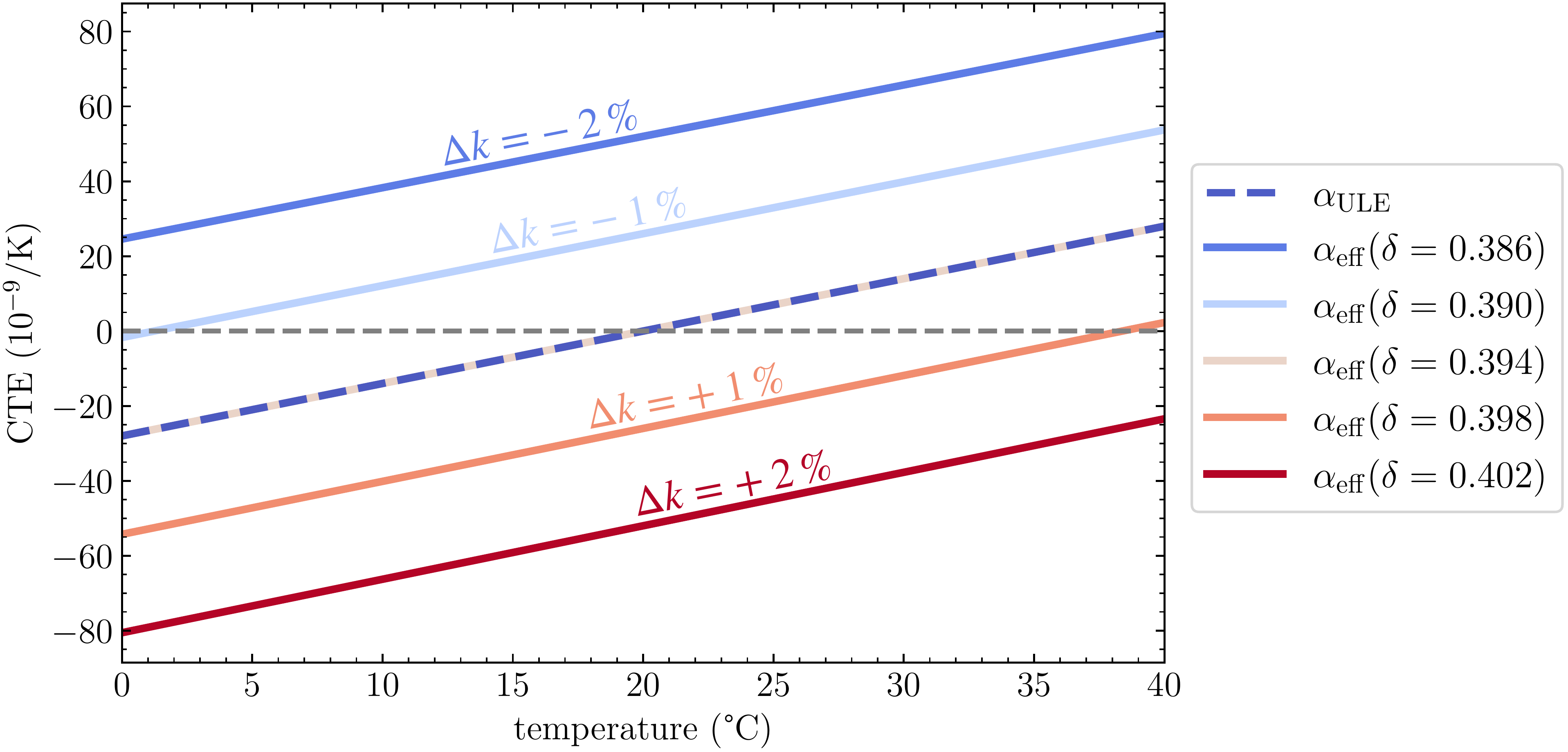}
    \caption{Effective CTE of a silicon--ULE resonator for $k=1$ and for small deviations from this value. For $k=1$, the effective CTE equals that of ULE, even though silicon is used as the spacer. Small variations in the correction factor shift the ZCT, while the slope remains nearly identical to that of ULE.}
    \label{fig:CTE_T_k=1}
\end{figure}

Figure~\ref{fig:CTE_T_k=1} shows the calculated effective CTE of such a resonator with $k=1$. Small deviations from this ideal correction factor have only a minor influence on the slope of the effective CTE, but introduce an offset leading to a shift in the ZCT.

The relatively large inner bore allows tuning of the coupling coefficient $\delta$ and thereby adjustment of $k$.
In practice, the resonator should first be designed such that $\delta$ is slightly below the value required for the target $k$.
For a fixed mirror geometry, the spacer provides several accessible tuning parameters. Both spacer length and inner bore diameter significantly influence the coupling coefficient.
However, reducing the spacer length is technically demanding, as the front surfaces must maintain high planarity and extremely low surface roughness. Likewise, modifying the inner bore diameter represents a nontrivial machining step.
We therefore propose adjusting the chamfer at the inner bore as a practical and effective method to increase $k$ toward the desired value.
Depending on the resonator geometry, the chamfer can increase $k$ by up to \SI{50}{\percent}.
It thus provides a convenient tuning parameter for the effective ZCT.
Increasing $k$ shifts $T_{\mathrm{eff}}$ toward higher temperatures and simultaneously reduces the slope of the effective CTE.

Although ULE is an excellent mirror substrate due to its ultra-low CTE, it ultimately limits the resonator’s frequency stability, as Brownian thermal noise from both the substrate and the mirror coatings becomes dominant~\cite{Kessler_2012}.
For the silicon–ULE design, a \SI{20}{\milli\meter}-long resonator would result in a fractional frequency instability of $\mathrm{mod}\,\sigma_y = \num{3.3e-15}$ (SiO$_2$/Ta$_2$O$_5$) or $\mathrm{mod}\,\sigma_y = \num{2.6e-15}$ (AlGaAs/GaAs).
For this estimate, we have used the same parameters as for the stability calculation presented in the previous section.
These limits correspond to the stability of currently available commercial resonator systems and meets the requirements of many technical applications.
Silicon-based resonators have demonstrated exceptionally low frequency drift rates~\cite{Lee2026}, making silicon a highly promising spacer material.

\subsection{Silicon resonators with fused silica mirrors at room temperature}

The final design concept corresponds to the general approach introduced in Eq.~(\ref{eq:general_k}). 
It enables resonator configurations that have previously not been considered viable. 
For instance, a combination of a silicon spacer with FS mirrors at room temperature is typically disregarded due to the comparatively large CTEs of both materials. 
In this section, we evaluate the performance of such a silicon–FS resonator optimized for room-temperature operation.

If the resonator is to be designed for a target ZCT of \SI{25}{\degreeCelsius}, the CTEs of silicon~\cite{Lyon1977} and fused silica (FS)~\cite{Jacobs1981} must be evaluated in the vicinity of this temperature. 
Since the temperature dependence of the CTE is generally nonlinear, the intrinsic material ZCTs cannot be directly used in this context. 
Instead, the CTEs are locally approximated by a linear expansion around the temperature range of interest. 
This linearization yields effective slopes and corresponding extrapolated ZCTs of the local approximation.

Using room-temperature coefficients of $a'_{\mathrm{s}}=\SI{8.9e-9}{\per\kelvin\squared}$ and an extrapolated $T'_{\mathrm{s}}=\SI{8}{\kelvin}$ for silicon, as well as $a'_{\mathrm{m}}=\SI{5.3e-9}{\per\kelvin\squared}$ and $T'_{\mathrm{m}}=\SI{185}{\kelvin}$ for fused silica~\cite{Jacobs1981}, we obtain a correction factor of $k=\num{1.3}$.
The resulting resonator has an effective CTE slope of $a_{\mathrm{eff}}=\SI{4.2e-9}{\per\kelvin\squared}$ in its ZCT at \SI{25}{\degreeCelsius}, which is only a factor of three higher than that of ULE glass, but a factor of two smaller than the slope obtained for cordierite~\cite{Ito2017}.
A \SI{20}{\milli\meter}-long silicon--FS resonator with the same parameters mentioned above would yield a predicted thermal noise floor of only $\mathrm{mod}\,\sigma_y = 2.2\times10^{-15}$ (SiO$_2$/Ta$_2$O$_5$) or $\mathrm{mod}\,\sigma_y = \num{7.8e-16}$ (AlGaAs/GaAs) at \SI{1}{\s} averaging time. 

A resonator based on this concept simultaneously provides several crucial advantages: a high Young's modulus, which reduces acceleration sensitivity; a low mechanical loss, which lowers thermal noise; a low linear drift rate, which improves long-term stability; and an ultra-low effective CTE, which enables excellent thermal stability.  
Such a combination of properties has not been achieved with conventional resonator designs without resorting to cryogenic operation.

The considerations above demonstrate that a silicon spacer with FS mirrors can indeed be optimized for room-temperature operation.
However, it must be checked whether the optical contact can withstand the forces that occur, as both materials have a relatively high CTE.
If necessary, the bonding procedure must be optimized to ensure a sufficient contact. 
In this context, advanced bonding techniques such as hydroxyl-bonding may provide a suitable alternative to conventional optical contacting.

A similar performance may also be achievable with other material combinations.
Materials such as diamond or sapphire are particularly attractive for future resonator generations, as they offer significantly higher Young's moduli than silicon, enabling reduced acceleration sensitivity.
Using mirror substrates such as diamond, silicon, or sapphire may, however, increase the resonator's thermal noise floor due to their higher CTEs, which lead to increased thermoelastic noise in the substrate contribution.

These results demonstrate that the presented approach enables compact optical resonators to reach performance levels that were previously attainable only under specialized conditions, such as cryogenic operation.
While the present work focuses on a room-temperature design, the underlying concept is fully applicable to other operating temperatures, including those relevant for space missions and other environments with nonstandard thermal conditions.

\section{Conclusion}

In this study, we investigated the CTE mismatch between cordierite spacers and ULE mirror substrates, motivated by the growing interest in cordierite as a spacer material for optical resonators. 
We provide key information on the coupling coefficient required to determine the individual CTE contributions of such resonators. 
We measured the CTE of a cordierite resonator with ULE mirrors and extracted the CTE parameters of cordierite, which have not previously been reported in literature. 
We find a CTE slope of \SI{8.811+-0.025e-9}{\per\K\squared}, significantly larger than that of ULE, resulting in small shifts of the effective ZCT when combining cordierite spacers with FS mirrors. 
This result suggests that compensation rings are not required for cordierite-based resonators. 
However, due to the increased temperature sensitivity, careful suppression of temperature fluctuations and gradients within the vacuum chamber remains essential.

We therefore proposed a zero-CTE resonator based on a cordierite spacer combined with ULE mirrors with a thermal sensitivity reduced by a factor of 18 compared to an all-ULE resonator.
This concept is particularly relevant for vacuum-gap cavities~\cite{Liu2024,McLemore2024}, which exhibit increased sensitivity to ambient temperature fluctuations.

We further extended our concept to two additional design approaches. 
First, we showed that a resonator can be designed such that its effective CTE is determined solely by the mirror substrate material, rendering the spacer material irrelevant for thermal expansion. 
Second, we introduced an innovative resonator concept that enables the use of silicon spacers with FS mirror substrates at room temperature while maintaining ultra-low thermal expansion.
The ideas outlined here can significantly improve the performance of compact resonators and pave the way for future practical applications of ultra-stable laser systems.

\begin{backmatter}

\bmsection{Funding}
Deutsche Forschungsgemeinschaft (EXC 2123/2 QuantumFrontiers—390837967); ERC CoG MightyMirrors (101170022)

\bmsection{Acknowledgment}
The authors would like to thank Noah Syring and Tobias Ohlendorf for preparing the resonator and performing the CTE measurement. 

\bmsection{Disclosures}
The authors declare no conflicts of interest.

\bmsection{Data availability}
Data underlying the results presented in this paper are not publicly available at this time but may be obtained from the authors upon reasonable request.
\end{backmatter}

\bibliography{bibliography}

\end{document}